\documentclass[useAMS,usenatbib]{mnras}

\usepackage{times}
\usepackage{color}
\usepackage{breakurl}
\usepackage{color}
\usepackage{mathrsfs}
\usepackage{multicol}
\usepackage{bm}		
\usepackage{pdflscape}	
\usepackage{amsfonts,amsmath,amssymb,mathrsfs}
\usepackage{graphicx,epsf,epsfig}
\usepackage{inputenc}
\usepackage{subfigure}
\usepackage{bm}
\usepackage{longtable}
\usepackage[usenames,dvipsnames]{xcolor}
\usepackage{multirow}
\usepackage{hyperref}
\usepackage{times}
\usepackage{color}
\usepackage{tabularx}
\usepackage{appendix}

\def\be{\begin{equation}}
\def\ee{\end{equation}}
\def\bea{\begin{eqnarray}}
\def\eea{\end{eqnarray}}

\title[]{Intermediate redshift calibration of Gamma-ray Bursts and cosmic constraints in non-flat cosmology}

\author[]{
Orlando Luongo$^{1,2,3}$\thanks{orlando.luongo@unicam.it}, Marco Muccino$^{3}$\thanks{marco.muccino@lnf.infn.it}
\\
$^{1}$Scuola di Scienze e Tecnologie, Universit\`a di Camerino, Via Madonna delle Carceri 9,  Camerino, 62032, Italy.\\
$^{2}$Dipartimento di Matematica, Universit\`a di Pisa, Largo Bruno Pontecorvo 5, Pisa, 56127, Italy.\\
$^{3}$NNLOT, Al-Farabi Kazakh National University, Al-Farabi av. 71,  Almaty, 050040, Kazakhstan.
}

\date{Accepted XXX. Received YYY; in original form ZZZ}

\pubyear{2022}

\begin{document}
\label{firstpage}
\pagerange{\pageref{firstpage}--\pageref{lastpage}}
\maketitle

\begin{abstract}
We propose how to calibrate long gamma-ray burst (GRB) correlations employing intermediate redshift data sets, instead of limiting to $z\simeq0$ catalogs. To do so, we examine the most updated observational Hubble data (OHD) and baryonic acoustic oscillations (BAO). We exploit the model-independent technique of \emph{B\'ezier polynomial interpolation}, alleviating \textit{de facto} the well-known circularity problem affecting GRB correlations. To get constraints on cosmic  parameters, using Markov chain Monte Carlo Metropolis algorithm, we distinguish the influence on BAO scale, $r_{\rm s}$, Hubble constant $H_0$, luminosity distance $D_{\rm L}(z)$ and spatial curvature $\Omega_k$. Inspired by the fact that a few 0.4$\%$ error on $r_{\rm s}$ is got from Planck results, utterly small compared with current BAO measurement errors, we discern two main cases, namely $(r_{\rm s}/r_{\rm s}^{\rm fid})=1$ and $(r_{\rm s}/r_{\rm s}^{\rm fid})\neq1$. For each occurrence, we first fix and then leave free the Universe's spatial curvature. In all our treatments, we make use of the well-consolidated \textit{Amati} correlation, furnishing tighter constraints on the mass density than previous literature. In particular, our findings turn out to be highly more compatible with those got, adopting the $\Lambda$CDM paradigm, with standard candle indicators. Finally, we critically re-examine the recent $H_0$ tension in view of our outcomes.
\end{abstract}


\section{Introduction}
\label{intro}

GRBs are often challenged as possible distance indicators and their use in cosmology is currently debated \citep{2021Galax...9...77L}. Their possible use is essential to highlight possible departures from the concordance $\Lambda$CDM paradigm, whose overall dynamics is described by six free parameters \citep{2021arXiv210505208P}, albeit at late times it can be well-approximated only by matter, $\Omega_m$, accounting for about the $30\%$ of the total energy budget \citep{Ratra1988}. The cosmological constant density drives the universe to accelerate and its experimentally and statistically agreement is robust, albeit recent tensions have been raised \citep{2021CQGra..38o3001D}, indicating mild discrepancies from the standard model predictions\footnote{For different perspectives about extensions of the standard model, see e.g. \citep{nostro,2022arXiv220402190D,mio2022}.} \citep{Sotiriou:2008rp}. 
Moreover, in the concordance paradigm, one conventionally assumes a \emph{perfectly spatially-flat} universe, supported by several observations\footnote{Cosmic microwave background  observations, for instance, seem to favor this fact, having  $\Omega_k=-0.001\pm0.002$ at $68\%$ confidence level. Further, inflationary paradigms also require severe limits on spatial curvature \citep{Tsujikawa2013}.}. 
In this sense, bounding spatial curvature even remains as an additional open caveat of cosmology \citep{2018ApJ...864...80O}. 

Intermediate and high redshifts data exceeding the redshift limits of supernovae Ia (SNe Ia) detectability, placed at $z\simeq 2.3$ \citep{Rodney2015}, and of other cosmic indicators, in general at $z\lesssim 3$,
turn out to be essential in order to shed light into the nature of those constituents pushing up the acceleration of the universe\footnote{Alleviating the severe difficulty to detect if the fluid responsible for the cosmic speed up is under the form of a pure cosmological constant or is a time-dependent dark energy one.} and to disclose whether the $\Lambda$CDM model may be seen as the final scenario describing large-scale dynamics or a limiting case of a more general landscape \citep{2019IJMPD..2830016C,2020arXiv200309341C}.

In this respect, GRBs could represent a plausible new class of cosmological indicators. These explosions are detectable up to $z=9.4$ \citep{Salvaterra2009,Tanvir2009,Cucchiara2011} and so attempts toward their use in cosmology as genuine cosmic indicators are currently highly debated. Essentially a way out to relate GRB photometric and spectroscopic properties is the missing puzzle piece  \citep{Amati2002,Ghirlanda04,Amati2008,Schaefer2007,CapozzielloIzzo2008,Dainotti2008,Bernardini2012,AmatiDellaValle2013,Wei2014,Izzo2015,Demianski17a,Demianski17b}. Moreover, the so-called \emph{circularity problem} arises, \textit{i.e.}, the calibration issue between radiated energy or luminosity and the spectral properties that becomes plausible only if a background cosmology is \emph{a priori} imposed\footnote{The calibration procedure is also debated. For a different perspective, see e.g. \citet{2021JCAP...09..042K}.}. Consequently, in view of the great uncertainty surrounding a model-independent procedure to calibrate GRBs, finding out new correlations that are background-independent becomes a crucial step to heal circularity. 

In this paper, we assume the widely-consolidate model-independent technique of GRB calibration constructed by means of B\'ezier polynomials \citep{LM2020} and we apply it to the $E_{\rm p}$--$E_{\rm iso}$ or \textit{Amati correlation} \citep[see e.g.,][]{Amati2008,AmatiDellaValle2013}, in particular to one of the best data set composed of $118$ GRBs \citep{2021JCAP...09..042K}.
We propose a calibration procedure involving intermediate redshift catalogs, in lieu of $z\simeq0$ data points, more often developed in the literature. We employ cosmic chronometers or OHD \citep[see][and references therein]{2018MNRAS.476.3924C}, and the most recent measurements of BAO \citep[see][and references therein]{2021MNRAS.504..300C}. To do so, we fit both OHD and BAO data in conjunction by means of two B\'ezier parametric curves. These approximated curves are interconnected since BAO measurements contain information on $H(z)$, without assuming an \textit{a priori} hypothesis on the universe spatial curvature. 
Through this calibration procedure, the above GRBs can be viewed as \emph{standardized objects} and can be used to test the standard spatially flat $\Lambda$CDM model and its minimal extension adding a non-zero spatial curvature parameter $\Omega_k$. 
We thus investigate the effects of $\Omega_k\neq0$ and fix constraints over the free parameters by means of Markov chain -- Monte Carlo (MCMC) analyses. The obtained results are not perfectly compatible with the current expectations since they roughly differ from Planck results on $\Omega_k$. We investigate the corresponding systematics and show tighter bounds over the mass density, quite more similar to those found using standard candles. Further, we demonstrate that both spatial curvature and $H_0$ tension cannot be easily fixed by GRBs, showing  larger values of $\Omega_m$ with respect to Planck results. Last but not least, the $H_0$ tension cannot be avoided even if $\Omega_k\neq0$.

The paper is divided into five main sections. In section \ref{sec:2}, we describe the main ingredients of our B\'ezier model-independent reconstructions. In section \ref{sec:3}, we work out our experimental results, including both calibration and MCMC outcomes. In section \ref{sec:4}, we theoretically interpret our findings and finally in section \ref{sec:5} we develop conclusions and perspectives of our work.

\section{Theoretical warm-up}
\label{sec:2}

The mostly-adopted and investigated GRB correlation in the literature is built up through the rest-frame peak energy $E_{\rm p}$ of the $\gamma$-ray time-integrated $\nu F_\nu$ energy spectrum and the isotropic energy $E_{\rm iso}$ radiated in $\gamma$-rays, \textit{i.e.},
\begin{equation}\label{eiso}
 E_{\rm iso}\equiv 4\pi D_{\rm L}^2 S_{\rm b}(1+z)^{-1}\,,   
\end{equation}
where the observed bolometric GRB fluence $S_{\rm b}$ is evaluated from the integral of the $\nu F_\nu$ spectrum in the rest-frame $1-10^4$~keV energy band. Finally, a correction factor, namely $(1+z)^{-1}$, is involved in order to take into account cosmological redshift effects, \textit{i.e.}, to transform the inferred GRB overall duration into the source cosmological rest-frame measurements.

In this respect, it appears obvious that using GRBs with the aim of fitting cosmic data may be strongly affected by a few uncertainties, caused by  selection and instrumental effects.

\subsection{Building up the correlation}

In view of the above considerations, the \textit{Amati} correlation \citep{Amati2002,Amati2008,AmatiDellaValle2013,Demianski17a,Dainotti18}, typically dubbed $E_{\rm p}-E_{\rm iso}$ relation, easily writes 
\begin{equation}
\label{Amatirel}
\log\left(\frac{E_{\rm p}}{{\rm keV}}\right)= a_0 + a_1 \log\left(\frac{E_{\rm iso}}{10^{52}{\rm erg}}\right)\,,
\end{equation}
where the functional dependence requires an intercept $a_0$ and a slope $a_1$. In addition, we need to fix the dispersion $\sigma_{\rm ex}$ \citep{Dago2005} and all the latter free terms require calibration.

We immediately see the caveat in Eq.~\eqref{eiso}, there $E_{\rm iso}$ depends on the background, \textit{i.e.}, the Hubble rate might be known \emph{a priori} and so the corresponding dependence on  the luminosity distance $D_{\rm L}$ is unavoidable.

\subsection{GRB data set}

To fulfill our fits, adopting the \textit{Amati} relation, we here employ the most recent and largest data set of $118$ bursts fulfilling the \textit{Amati} correlation itself. For these data points, we underline they provide the smallest intrinsic dispersion \citep{2021JCAP...09..042K}.

Moreover, to get model-independent cosmological bounds, in the following we now need to calibrate the relation by means of model-independent techniques, as below reported. Before that, we introduce the intermediate redshift data sets through which we intend to calibrate our GRB data set.

\subsection{Intermediate redshift data sets}

To calibrate GRB data sets we employ 
\begin{itemize}
\item[--] {\bf OHD}, consisting at present of $31$ measurements of the Hubble rate $H(z)$. The corresponding measures are got at different redshifts \citep[see, e.g.,][]{2018MNRAS.476.3924C}.
\item[--] {\bf BAO}, consisting of $15$ measurements, split into $9$ uncorrelated and $6$ correlated, see Table~\ref{tab:BAO}.
\end{itemize}
\begin{table}
\centering
\setlength{\tabcolsep}{.8em}
\renewcommand{\arraystretch}{1.}
\begin{tabular}{lcccc}
\hline\hline
Survey          & $z$       &   $\delta$            
&  $r$                      
& Ref.\\
                &           &  [Mpc]                &   &   \\
\hline
6dFGS	        & $0.097$   &   $372^{+115}_{-50}$	
&  $0.997^{+0.005}_{-0.004}$      
&   [1]\\
SDSS MGS	    & $0.150$    &	$664^{+25}_{-25}$
&  $0.989^{+0.003}_{-0.003}$
& [2]\\
SDSS DR7	    & $0.275$   &   $1104^{+30}_{-30}$
&  $0.958^{+0.002}_{-0.002}$
& [3]\\
BOSS DR11	    & $0.320$	&   $1264^{+25}_{-25}$		
&  $0.985^{+0.001}_{-0.001}$
& [4]\\
SDSS DR7 LRG    & $0.350$	&   $1356^{+25}_{-25}$
&  $0.963^{+0.005}_{-0.005}$
& [5]\\
BOSS DR11	    & $0.570$	&   $2056^{+20}_{-20}$
&  $0.985^{+0.002}_{-0.002}$
& [4]\\
eBOSS DR14 LRG	& $0.720$	&   $2377^{+61}_{-59}$
&  $0.995^{+0.002}_{-0.002}$
& [6]\\
eBOSS DR14	    & $1.520$	&   $3843^{+147}_{-147}$
&  $0.995^{+0.002}_{-0.002}$
& [7]\\
eBOSS DR16	    & $2.334$	&   $4549^{+96}_{-96}$
&  $0.998^{+0.003}_{-0.003}$
& [8]\\
\hline
WiggleZ		    & $0.44$	&   $1716^{+83}_{83}$
&  $0.990^{+0.002}_{-0.002}$
& [9]\\
WiggleZ		    & $0.60$     &	$2221^{+101}_{-101}$
&  $0.990^{+0.002}_{-0.002}$
& [9]\\
WiggleZ		    & $0.73$	&   $2516^{+86}_{-86}$
&  $0.990^{+0.002}_{-0.002}$
& [9]\\
BOSS DR12	    & $0.38$    &   $1477^{+16}_{-16}$
&  $0.995^{+0.002}_{-0.002}$
& [10]\\
BOSS DR12	    & $0.51$	&   $1877^{+19}_{-19}$
&  $0.995^{+0.002}_{-0.002}$
& [10]\\
BOSS DR12	    & $0.61$	&   $2140^{+22}_{-22}$
&  $0.995^{+0.002}_{-0.002}$
& [10]\\
\hline
\end{tabular}
\caption{List of the redshift $z$, the $\delta$ measurement and the ratio $r$ of the BAO data considered in this work. The upper part of the Table consists of $9$ uncorrelated measurements, whereas the lower part displays $6$ correlated ones. References: [1] \citet{2018MNRAS.481.2371C}, [2] \citet{Aubourg15}, [3] \citet{Percival10}, [4] \citet{2014MNRAS.441...24A}, [5] \citet{2012MNRAS.427.2132P}, [6] \citet{2018ApJ...863..110B}, [7] \citet{2018MNRAS.473.4773A}, [8] \citet{2020ApJ...901..153D}, [9] \citet{2014MNRAS.441.3524K}, [10] \citet{2017MNRAS.470.2617A}.}
\label{tab:BAO} 
\end{table}

For the sake of clearness, we focus on BAO data. The latter  are often provided as volume-averaged distances $D_{\rm V}^{\rm obs}=\delta(r_{\rm s}/r_{\rm s}^{\rm fid})$, where $r_{\rm s}$ is the comoving sound horizon at the baryon-drag epoch and $r_{\rm s}^{\rm fid}$ is the value of $r_{\rm s}$ for the fiducial cosmological model used to convert redshift to distances.
For each measurement, $\delta$ and
\begin{equation}\label{theratio}
r\equiv(r_{\rm s}/r_{\rm s}^{\rm fid})\,,  
\end{equation}
are given in the third and fourth columns of Table~\ref{tab:BAO}, respectively.

The ratio $r$ has been computed by using the fiducial values $r_{\rm s}^{\rm fid}$ provided in the references listed in Table~\ref{tab:BAO}. In particular,  the value of $r_{\rm s}$ is taken from \citet{Planck2018}.

\noindent Hence, by construction, BAO measurements provide
\begin{equation}
\label{DVth}
    D_{\rm V}(z)=\left[\frac{c\,z}{(1+z)^2}\frac{D_{\rm L}^2(z)}{H(z)}\right]^{1/3}\,,
\end{equation}
showing constraints on the cosmological parameters through their influence on the combined action of $r_{\rm s}$, $H(z)$ and $D_{\rm L}(z)$.

For the standard model\footnote{For the sake of clearness, we can state \emph{for standard models}, involving the $\Lambda$CDM paradigm and the Chevallier-Polarski-Linder parametrization.}, the 0.4$\%$ error on $r_{\rm s}$ from Planck results \citep{Planck2018} is small compared to current BAO measurement errors, therefore the constraints come mainly through $H(z)$ and $D_{\rm L}(z)$ \citep{Aubourg15}.

To check the influence of $r$ on the estimate of the cosmological parameters, we consider both $r=1$ and $r\neq1$ cases.

\subsection{Statistical errors}

In principle, it is worth noticing that to reduce the statistical errors, one could adopt since the beginning  the larger and most recent SNe Ia catalog. Such data points are prompted either under the form of a catalog of distance moduli $\mu_{\rm SN}(z)$ (related to $D_{\rm L}$, see, e.g., \citealt{2018ApJ...859..101S}) or as $E(z)\equiv H(z)/H_0$ data set \citep{2018ApJ...853..126R}, where $H_0$ is the Hubble constant, albeit in the case of null spatial curvature. However, the inclusion of SN Ia data has two main drawbacks, below summarized. 
\begin{itemize}
    \item[--] As a first possibility, luminosity distance measurements from the definition of $\mu_{\rm SN}(z)$ could have been used instead of BAO data and in conjunction with OHD. However, this joint analysis can be done only if one assumes that the spatial curvature of the Universe is zero \citep[see, e.g.,][for details]{2019MNRAS.486L..46A}.
    \item[--] As a second possibility, $E(z)$ from SNe Ia could have been employed together with OHD to constrain $H(z)$ and extract $D_{\rm L}(z)$ from BAO, as shown in Eq.~\eqref{DVth}. However, again, $E(z)$ from SNe Ia have been established by assuming a flat spatial curvature.
\end{itemize}
Hence, as shown in Eq.~\eqref{DVth}, the only possibility to obtain constraints on $D_{\rm L}(z)$ without imposing a \emph{a priori} spatial curvature consists in using only BAO data points in conjunction with OHD. We will follow this more reliable strategy in what follows.

\section{Model-independent calibrations of GRBs}
\label{sec:3}

Bearing in mind all the above ingredients, our prescription resides in interpolating OHD and BAO data sets without acquiring any \emph{cosmological model}, namely to involve a model-independent calibration technique\footnote{The words \emph{model-independent} typically rely on expansions and/or reconstructions of cosmic quantities, \textit{i.e.}, without any need of postulating a cosmological model, see e.g. \citep{Aviles:2012ay,Dunsby:2015ers}.}. The here-employed method is  attained by working out the so-called \emph{B\'ezier parametric curves}, obtained as a linear combination of Bernstein basis polynomials and firstly proposed in GRB contexts in \citet{2019MNRAS.486L..46A}. The main steps to follow are thus summarized below.

\subsection{B\'ezier polynomials and GRB calibration}

The most general B\'ezier curve with established order $n$, constructed by means of OHD data, can be written by 
\begin{equation}
\label{bezier1}
H_n(x)=\sum_{i=0}^{n} g_\alpha\alpha_i h_n^d(x)\quad,\quad h_n^i(x)\equiv n!\frac{x^i}{i!} \frac{\left(1-x\right)^{n-i}}{(n-i)!}\,,
\end{equation}
where $\alpha_i$ are the coefficients of the linear combination of the polynomials $h_n^i(x)$, re-scaled by a conventional factor $g_\alpha=100$~km/s/Mpc, being  positive-defined for $0\leq x\equiv z/z_{\rm OHD}^{\rm max}\leq1$, with $z_{\rm OHD}^{\rm max}$ representing the maximum redshift of the OHD catalog.
As proved in \citet{LM2020}, the only possible non-linear monotonic growing function over the redshift range of OHD has order $n=2$, \textit{i.e.}, $H_2(z)$.
Moreover, by construction, it is possible to identify $\alpha_0$ with $h_0=H_0/(100$~km/s/Mpc$)$.

BAO data $\delta$ can be fitted by using Eq.~\eqref{DVth} and the function $H(z)$ can be approximated with $H_2(z)$, extrapolated up to the BAO maximum redshift $z_{\rm BAO}^{\rm max}$.
In so doing, it is therefore licit to use BAO measurements in order to fit $D_{\rm L}^2(z)$ in a \emph{cosmology-independent way}, again, by resorting a B\'ezier curve of order $m$
\begin{equation}
\label{bezier2}
D^2_m(y)=\sum_{j=0}^{m} g_\beta \beta_j d_m^j(y)\ \ ,\ \ d_m^j(y)\equiv m!\frac{y^j}{j!} \frac{\left(1-y\right)^{m-j}}{(m-j)!}\,,
\end{equation}
where $\beta_j$ are the coefficients of the linear combination of the polynomials $d_m^j(y)$ rescaled by a factor $g_\beta=1$~Gpc$^2$ and positive-defined for $0\leq y\equiv z/z_{\rm BAO}^{\rm max}\leq1$. Immediately, from the above construction one argues how to extend the use of B\'ezier curves to BAO data points and, so, one highlights that:
\begin{itemize}
    \item[--] by definition of cosmic distance, we need that  $D^2_m(0)\equiv0$ and so we are forced to start our B\'ezier expansion with $j=1$; 
    \item[--]the only non-linear monotonic growing function with the redshift has order $m=3$, \textit{i.e.}, $D^2_{13}(z)$ since $j$ runs from $1$ to $3$. 
\end{itemize}

Last but not least, it is remarkable  to stress that the above determination of the luminosity distance is quite general. In other words, it includes information on the Universe's spatial curvature, without introducing any theoretical bias, jeopardizing the overall picture and leading to circularity. 

We are now ready to calibrate our free coefficients and to introduce the statical methods of numerical analyses, as we prompt below. 

\subsection{Calibrating the coefficients with nested likelihoods}

We now estimate the coefficients $\alpha_i$ ($0\leq i\leq2$) and $\beta_j$ ($1\leq j\leq3$) through a nested likelihood approach. This method combines 
\vspace{0.3cm}
\begin{itemize}
    \item[1)] a fit of the OHD data in the range $0\leq z\leq z_{\rm OHD}^{\rm max}$, and 
\vspace{0.3cm}    
\item[2)] a fit of the BAO data in the range $0\leq z\leq z_{\rm BAO}^{\rm max}$, using also the extrapolation of $H_2(z)$ up to $z_{\rm BAO}^{\rm max}$ by defining the function
    \begin{equation}
        \label{bezier3}
        \Delta_{\rm V}(z)=\left[\frac{c\,z}{(1+z)^2}\frac{D_{13}^2(z)}{H_2(z)}\right]^{1/3}\,.
    \end{equation}
\end{itemize}
Assuming Gaussian distributed errors, the total log-likelihood function of the model-independent estimate of $D_{\rm L}(z)$ is given by
\begin{equation}
    \ln \mathcal{L}_{\rm D} = \ln \mathcal{L}_{\rm O} + \ln \mathcal{L}_{\rm B}\,, 
\end{equation}
where each contribution is described in details below.
\vspace{0.3cm}
\begin{itemize}
\item[--] 
For OHD the log-likelihood function reads as
\begin{equation}
\label{loglikeOHD}
    \ln \mathcal{L}_{\rm O} = -\frac{1}{2}\sum_{k=1}^{N_{\rm O}}\left\{\left[\dfrac{H_k-H_2(z_k)}{\sigma H_k}\right]^2 + \ln(2\pi\,\sigma H_k^2)\right\}\,,
\end{equation}
where $N_{\rm O}$ is the size of the OHD catalog with values $H_k$ and attached errors $\sigma H_k$.
\vspace{0.3cm}
\item[--]
For BAO the log-likelihood function is given by
\begin{equation}
\label{loglikeBAO}
\ln \mathcal{L}_{\rm B} = -\frac{1}{2}\sum_{k=1}^{N_{\rm B}} \left\{\left[\frac{\delta_k-\Delta_{\rm V}(z_k)}{\sigma\delta_k}\right]^2 + \ln(2\pi\,\sigma\delta_k^2)\right\}\,,
\end{equation}
where $N_{\rm B}$ is the size of the BAO catalog with values $\delta_k$ and attached errors $\sigma\delta_k$.
\end{itemize}

\begin{figure*}
\centering
\includegraphics[width=0.49\hsize,clip]{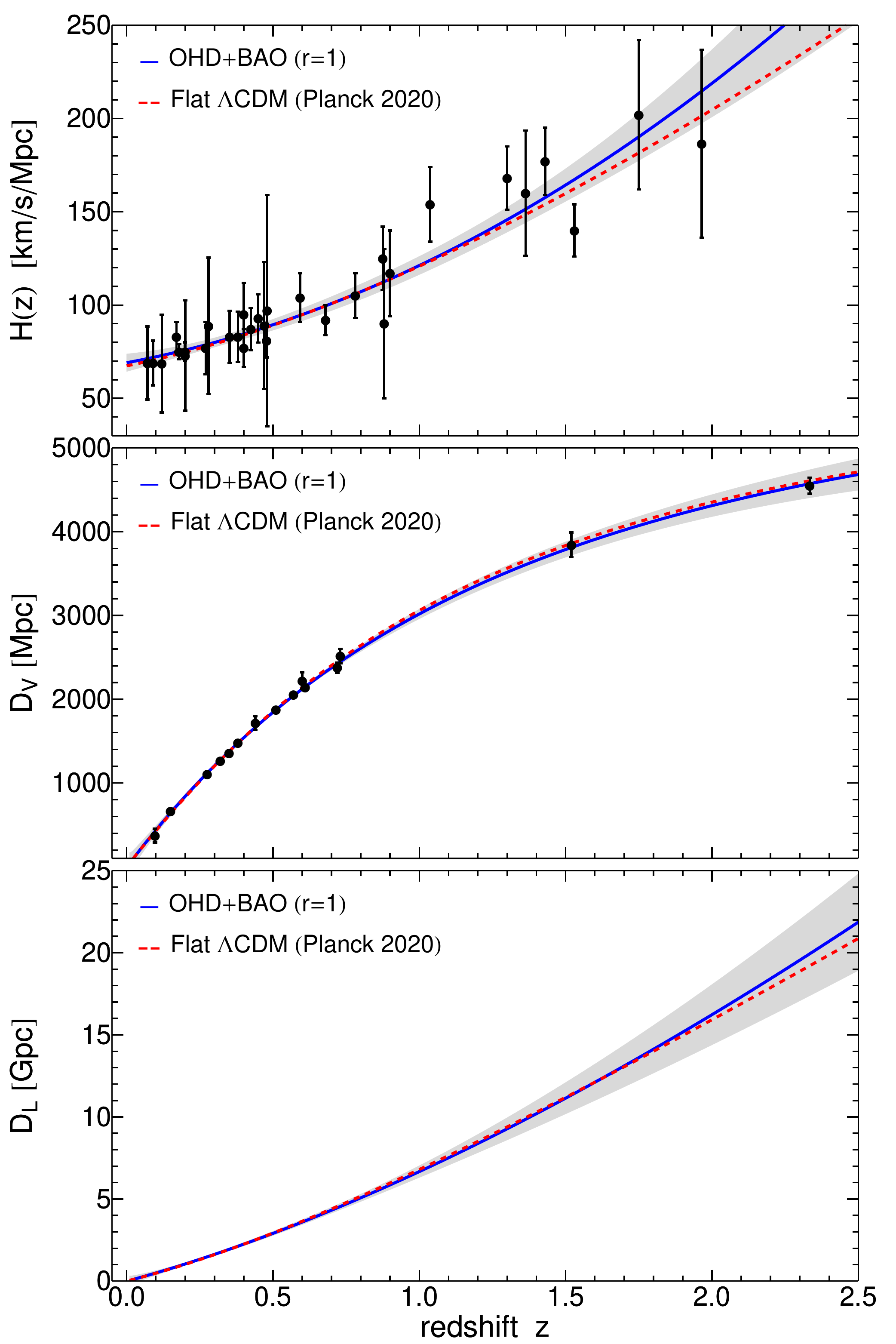}
\includegraphics[width=0.49\hsize,clip]{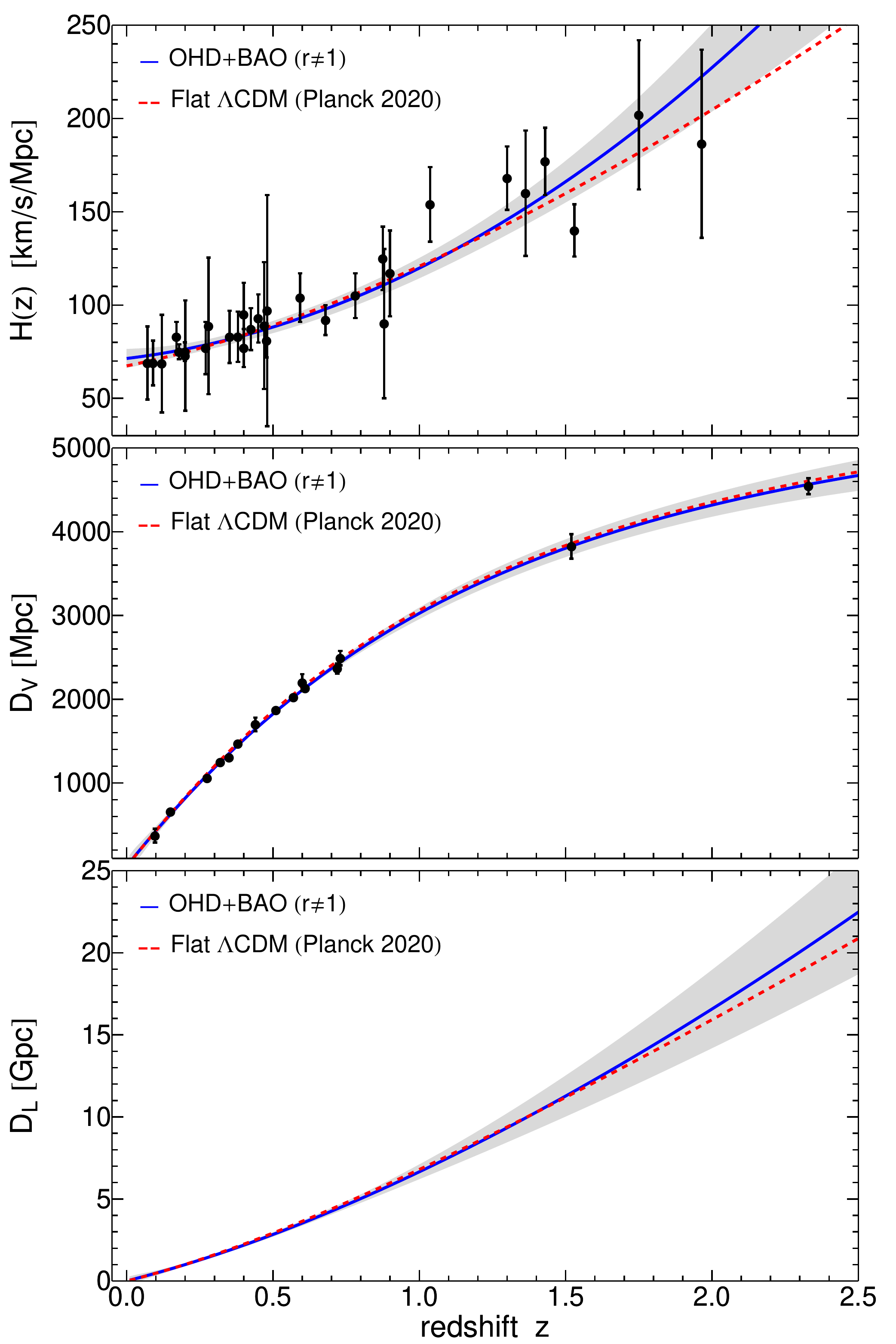}
\caption{Plots of the best-fitting B\'ezier curves (blue thick lines) approximating $H(z)$, $D_{\rm V}(z)$, and $D_{\rm L}(z)$ with the $1$--$\sigma$ confidence bands (gray shaded areas): \textit{left} for $r=1$ and \textit{right} for $r\neq1$. A comparison with the $\Lambda$CDM paradigm \citep{Planck2018} is also shown (see dashed red curves).}
\label{fig:Bez}
\end{figure*}

\begin{figure*}
\centering
\includegraphics[width=0.49\hsize,clip]{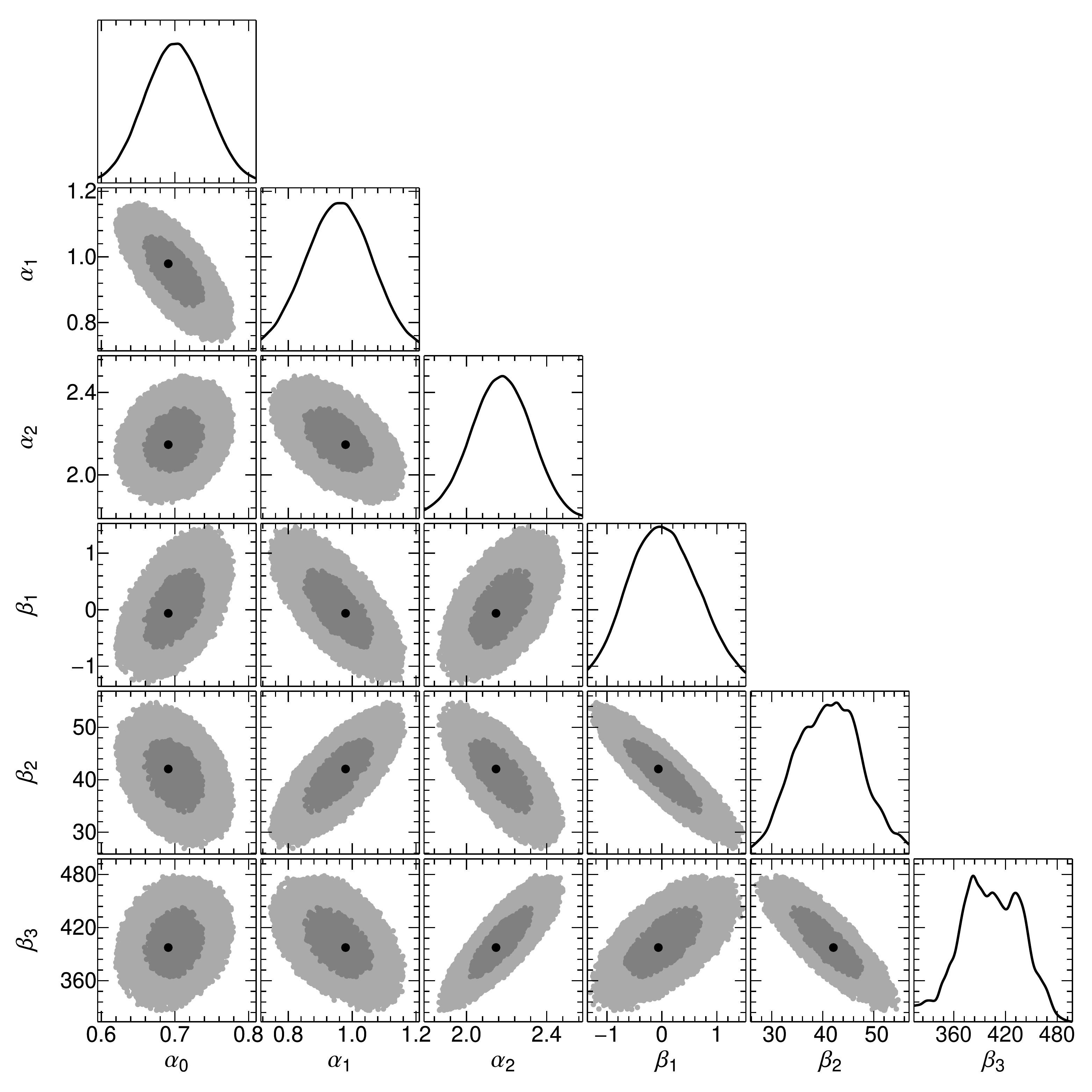}
\includegraphics[width=0.49\hsize,clip]{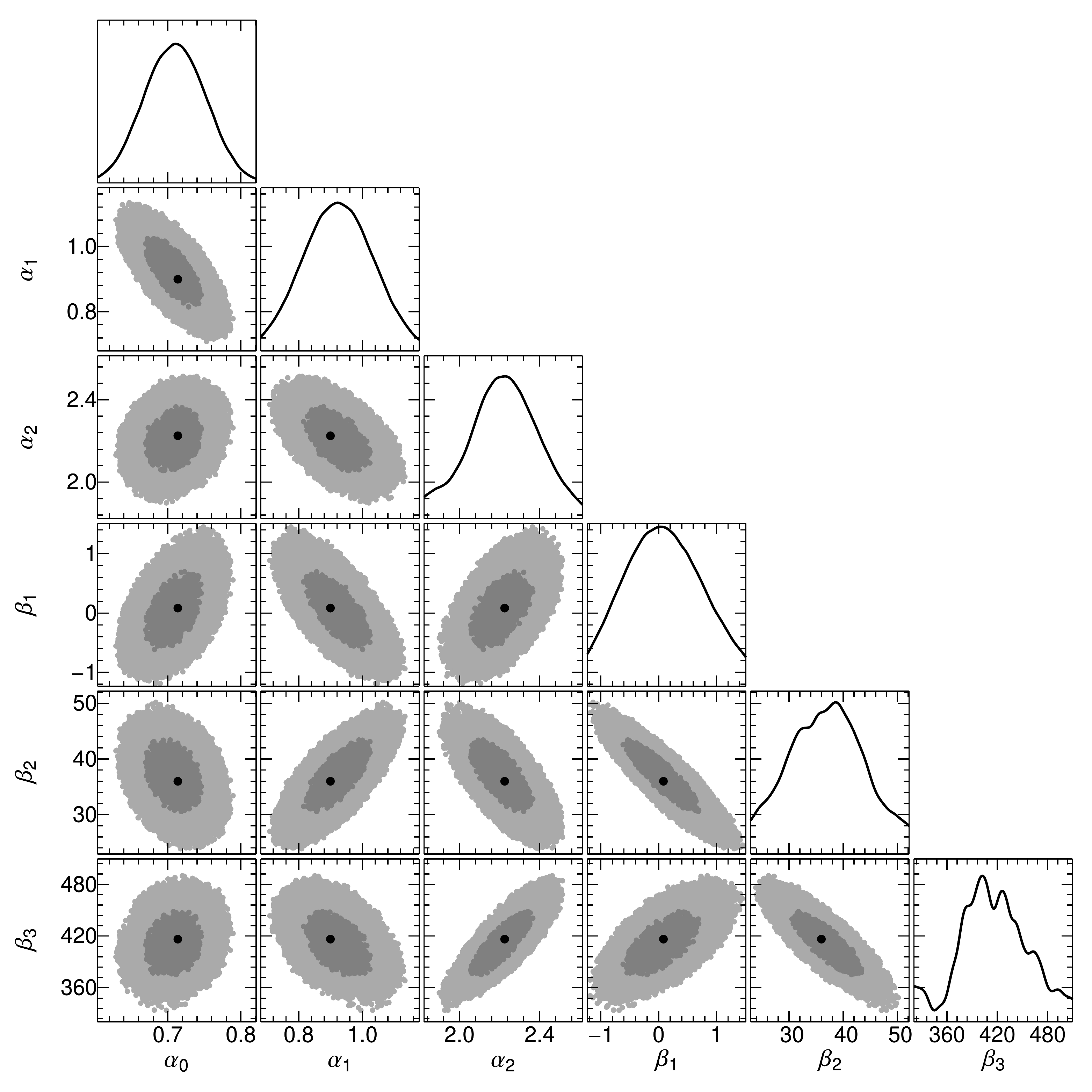}
\caption{Contour plots of the best-fit B\'ezier  coefficients $\alpha_i$ of $H_2(z)$ and $\beta_j$ of $D^2_{13}(z)$: \textit{left} for $r=1$ and \textit{right} for $r\neq1$. Darker (lighter) regions mark the $1$--$\sigma$ ($2$--$\sigma$) confidence regions.}
\label{fig:Bez_cont}
\end{figure*}

The best-fit curves approximating both OHD and BAO catalogs and the resulting trend of the luminosity distance are portrayed in Fig.~\ref{fig:Bez}, where a comparison with the predictions of the $\Lambda$CDM paradigm \citep{Planck2018} are also shown.
The best-fit coefficients $\alpha_i$ and $\beta_j$, on which the plots in Fig.~\ref{fig:Bez} are based, are displayed in the contour plots of Fig.~\ref{fig:Bez_cont} and summarized in Table~\ref{tab:Bez}.

\begin{table*}
\centering
\setlength{\tabcolsep}{1.3em}
\renewcommand{\arraystretch}{.9}
\begin{tabular}{lcccccccc}
\hline\hline
                &                   &                   &   OHD             &     
                &                   &                   & BAO               &   \\
\hline
                &                   &$\alpha_0$         & $\alpha_1$        
                & $\alpha_2$        & \vline            & $\beta_1$         
                & $\beta_2$         & $\beta_3$   \\
\hline
$r=1$                  &\vline&
$0.691^{+0.057}_{-0.032}$   & 
$0.979^{+0.082}_{-0.128}$   & 
$2.147^{+0.173}_{-0.127}$   & 
\vline                      &
$-0.064^{+0.759}_{-0.594}$  & 
$42.071^{+5.511}_{-8.187}$  & 
$397.469^{+44.913}_{-32.815}$  \\
$r\neq1$                  &\vline&
$0.714^{+0.032}_{-0.044}$   & 
$0.899^{+0.124}_{-0.084}$   & 
$2.225^{+0.140}_{-0.169}$   & 
\vline                      &
$0.083^{+0.610}_{-0.698}$  & 
$35.984^{+7.420}_{-5.432}$  & 
$416.395^{+31.966}_{-41.151}$  \\
\hline
\hline
\end{tabular}
\caption{Best-fit B\'ezier coefficients $\alpha_i$ of $H_2(z)$ and $\beta_j$ of $D^2_{13}(z)$ for $r=1$ and $r\neq1$.}
\label{tab:Bez} 
\end{table*}

Next, to get numerical bounds on the cosmological parameters, we employ the reconstructed luminosity distance portrayed in Fig.~\ref{fig:Bez} to calibrate the \textit{Amati} correlation in a model-independent as 
\begin{align}
\label{Eisocal}
E_{\rm iso}^{\rm cal}(z) &\equiv 4\pi D_{13}^2(z) S_{\rm b}(1+z)^{-1}\,,\\
\label{sEisocal}
\sigma E_{\rm iso}^{\rm cal}(z) &\equiv E_{\rm iso}^{\rm cal}(z)\sqrt{\left[\frac{2\sigma D_{13}(z)}{D_{13}(z)}\right]^2 + \left(\frac{\sigma S_{\rm b}}{S_{\rm b}}\right)^2}\,,
\end{align}
where the error $\sigma E_{\rm iso}^{\rm cal}$ depends upon the errors on the GRB observable $S_{\rm b}$ and the reconstructed luminosity distance $D_{13}(z)$.

To successfully calibrate all GRBs in each catalog, one needs to extrapolate $D_{13}(z)$ at redshifts higher than those of OHD and BAO catalogs. 
This, in principle may add further bias in the estimate of the cosmological parameters.
To check this possibility, we performed a nested likelihood approach \citep{LM2020} that combines two sub-models involving: 
\vspace{0.3cm}
\begin{itemize}
    \item[i)]  a \textit{calibrator sample} composed of GRBs in the range $0\leq z\leq z^{\rm max}_{\rm BAO}$ (encompassing both OHD and BAO observations), employed for estimating the GRB correlation parameters, and 
    \vspace{0.3cm}
    \item[ii)] a \textit{cosmological sample}, i.e., the whole GRB data set, used to estimate the free model parameters.
\end{itemize}

Again, assuming Gaussian distributed errors, the total GRB log-likelihood function is given by
\begin{equation}
\label{a0}
    \ln \mathcal{L}_{\rm G} = \ln \mathcal{L}_{\rm G}^{\rm cal} + \ln \mathcal{L}_{\rm G}^{\rm cos}\,.
\end{equation}
The calibration log-likelihood is given by
\begin{equation}
\label{a1}
\ln \mathcal{L}_{\rm G}^{\rm cal} = -\frac{1}{2}\sum_{k=1}^{N_{\rm cal}}\left\{\left[\dfrac{Y_k-Y(z_k)}{\sigma Y_k}\right]^2 + \ln(2\pi\,\sigma Y_k^2)\right\}\,,\\
\end{equation}
where $N_{\rm cal}=65$ and
\begin{subequations}
\begin{align}
Y_{\rm k} \equiv&\, \log E_{{\rm p},k}\,,\\
\label{a2}
Y(z_k)\equiv&\, a_0 + a_1 \left[E_{\rm iso}^{\rm cal}(z_k)-52\right]\,,\\
\sigma Y_k^2 \equiv&\, \left(\sigma\log E_{{\rm p},k}\right)^2 + a_1^2\left[\sigma\log E_{\rm iso}^{\rm cal}(z_k)\right]^2+\sigma_{\rm ex}^2\,.
\end{align}
\end{subequations}
The cosmological log-likelihood is given by
\begin{equation}
\label{a4}
\ln \mathcal{L}_{\rm G}^{\rm cos} = -\frac{1}{2}\sum_{k=1}^{N_{\rm cos}}\left\{\left[\dfrac{\mu_k-\mu_{\rm th}(z_k)}{\sigma\mu_k}\right]^2 + \ln (2 \pi \,\sigma\mu_k^2)\right\}\,, 
\end{equation}
where we have $N_{\rm cos}=118$ and
\begin{subequations}
\begin{align}
\label{a5}
\mu_k\equiv& \,\frac{5}{2 a_1}\left[\log E_{{\rm p},k} - a_0 - a_1\log\left(\frac{4\pi S_{{\rm b},k}}{1+z_k}\right)\right]\,,\\
\sigma\mu_k^2 \equiv& \,\frac{25}{4 a_1^2}\left[{(\sigma\log E_{{\rm p},k}})^2 + a_1^2 (\sigma\log S_{{\rm b},k})^2+\sigma_{\rm ex}^2\right]\,.
\end{align}
\end{subequations}

\subsection{Cosmic background and bounds over B\'ezier fits}

We are now in position to experimentally test the non-flat $\Lambda$CDM paradigm by feeding it within $\mu_{\rm th}(z)$ from Eq.~\eqref{a4}. 
The corresponding Hubble rate becomes 
\begin{equation}
\label{acca}
H(z)= H_0\sqrt{\Omega_{\rm m}(1+z)^3+\Omega_{\rm k} (1+z)^2+\Omega_\Lambda}\,,
\end{equation}
where $\Omega_{\rm m}$, $\Omega_{\rm k}$ and  $\Omega_\Lambda=1-\Omega_{\rm m}-\Omega_{\rm k}$ are matter, curvature and cosmological constant density parameters, respectively. 
The luminosity distance \citep[see, e.g.,][]{Goobar1995} is
\begin{equation}
\label{dlHz}
D_{\rm L}(z)=\frac{c}{H_0}\dfrac{(1+z)}{\sqrt{|\Omega_k|}}S_k\left[\sqrt{|\Omega_k|}\int_0^z\dfrac{H_0 dz'}{H(z')}\right] ,
\end{equation}
where $S_k(x)=\sinh(x)$ for $\Omega_k>0$, $S_k(x)=x$ for $\Omega_k = 0$, and $S_k(x)=\sin(x)$ for $\Omega_k<0$, and the distance modulus is
\begin{equation}
\label{muz}
\mu_{\rm th}(z)= 25+5\log\left[\frac{D_{\rm L}(z)}{{\rm Mpc}}\right]\,.
\end{equation}

Finally, we perform MCMC fittings by searching for the best-fit parameters maximizing the log-likelihood defined in Eq.~\eqref{a0} and the $1$ and $2$--$\sigma$ contours. To do so, we modified the \texttt{Wolfram  Mathematica} code from \citet{2019PhRvD..99d3516A}.

Depending on $r$, we decided to fix $h_0$ with the values of $\alpha_0$ obtained from the B\'ezier fitting, namely, 
\begin{subequations}
\begin{align}\label{h0results}
&h_0=0.691^{+0.057}_{-0.032},\quad \text{for}\,\,\, r=1,\\  
&h_0=0.714^{+0.032}_{-0.044},\quad  \text{for}\,\,\, r\neq1.
\end{align}
\end{subequations}

So, we finally explore both cases with free $\Omega_{\rm k}$ and $\Omega_{\rm k}\equiv0$
The results are summarized in Table~\ref{tab:summarymodelslike}.
\begin{table*}
\centering
\setlength{\tabcolsep}{0.5em}
\renewcommand{\arraystretch}{1.5}
\begin{tabular}{lccccc}
\hline\hline
                &  $a_0$
                &  $a_1$
                &  $\sigma_{\rm ex}$                                    
                &  $\Omega_{\rm m}$
                &  $\Omega_{\rm k}$\\
\hline
$r=1$      
&  $1.82_{-0.06\,(-0.11)}^{+0.05\,(+0.10)}$
&  $0.70_{-0.03\,(-0.06)}^{+0.03\,(+0.07)}$
&  $0.28_{-0.01\,(-0.03)}^{+0.02\,(+0.04)}$
&  $0.25_{-0.09\,(-0.22)}^{+0.09\,(+0.21)}$
&  $0.25_{-0.31\,(-0.52)}^{+0.64\,(+1.57)}$\\
&   $1.81_{-0.05\,(-0.11)}^{+0.05\,(+0.10)}$
&  $0.71_{-0.03\,(-0.06)}^{+0.03\,(+0.07)}$
&  $0.29_{-0.02\,(-0.04)}^{+0.02\,(+0.04)}$
&  $0.27_{-0.06\,(-0.12)}^{+0.09\,(+0.20)}$
&  $0$\\              
\hline
$r\neq1$   
&  $1.82_{-0.05\,(-0.11)}^{+0.05\,(+0.10)}$
&  $0.70_{-0.03\,(-0.06)}^{+0.03\,(+0.07)}$
&  $0.28_{-0.02\,(-0.03)}^{+0.02\,(+0.04)}$
&  $0.21_{-0.07\,(-0.18)}^{+0.08\,(+0.18)}$
&  $0.34_{-0.35\,(-0.61)}^{+0.44\,(+1.25)}$\\
&  $1.82_{-0.05\,(-0.11)}^{+0.05\,(+0.10)}$
&  $0.71_{-0.03\,(-0.07)}^{+0.03\,(+0.07)}$
&  $0.29_{-0.02\,(-0.03)}^{+0.02\,(+0.04)}$
&  $0.26_{-0.06\,(-0.12)}^{+0.08\,(+0.18)}$
&  0\\
\hline
\end{tabular}
\caption{Nested log-likelihood best-fit results and $1$--$\sigma$ ($2$--$\sigma$) errors for flat and non-flat $\Lambda$CDM models and for $r=1$ and $r\neq1$.}
\label{tab:summarymodelslike}
\end{table*}
\begin{figure*}
\centering
\includegraphics[width=0.49\hsize,clip]{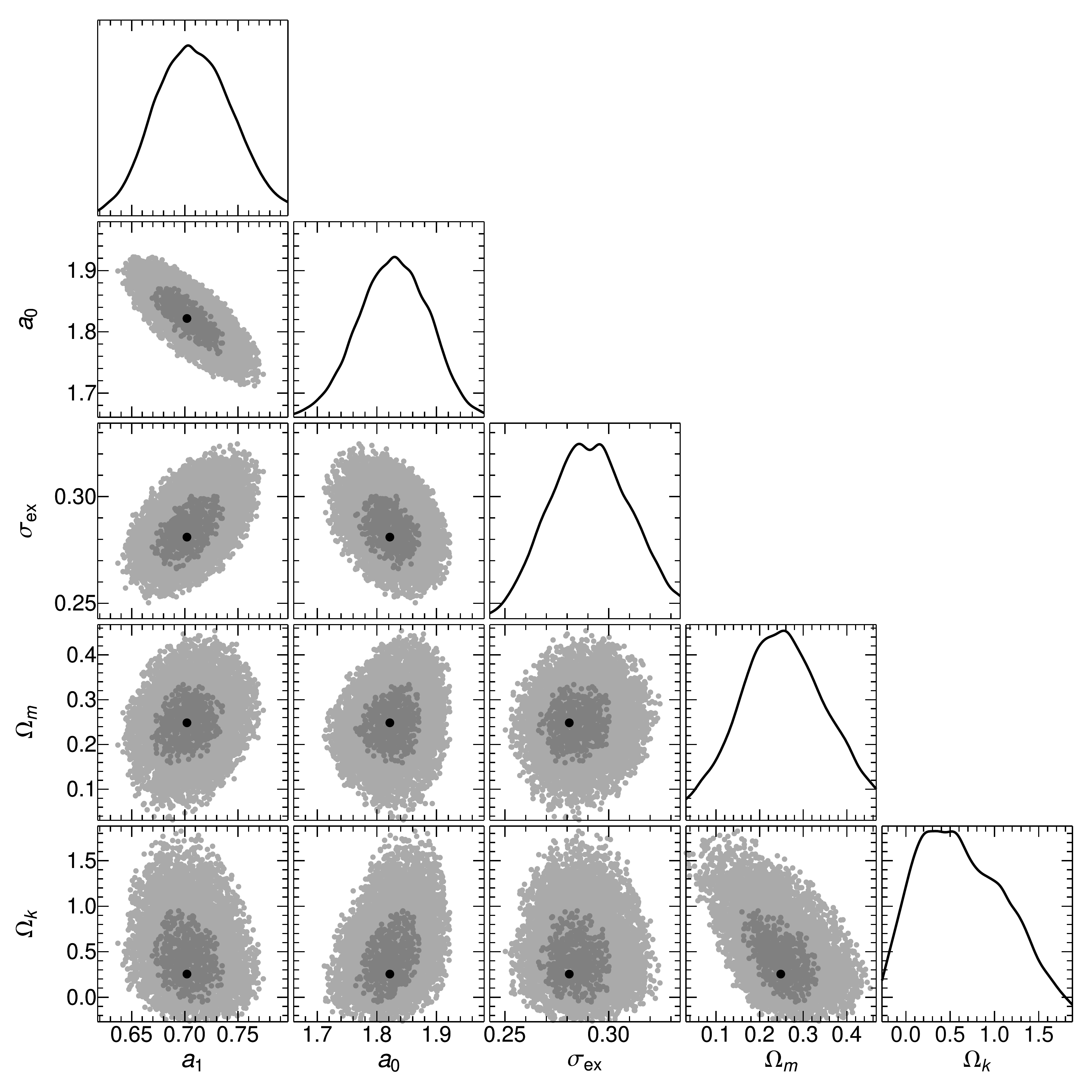}
\includegraphics[width=0.49\hsize,clip]{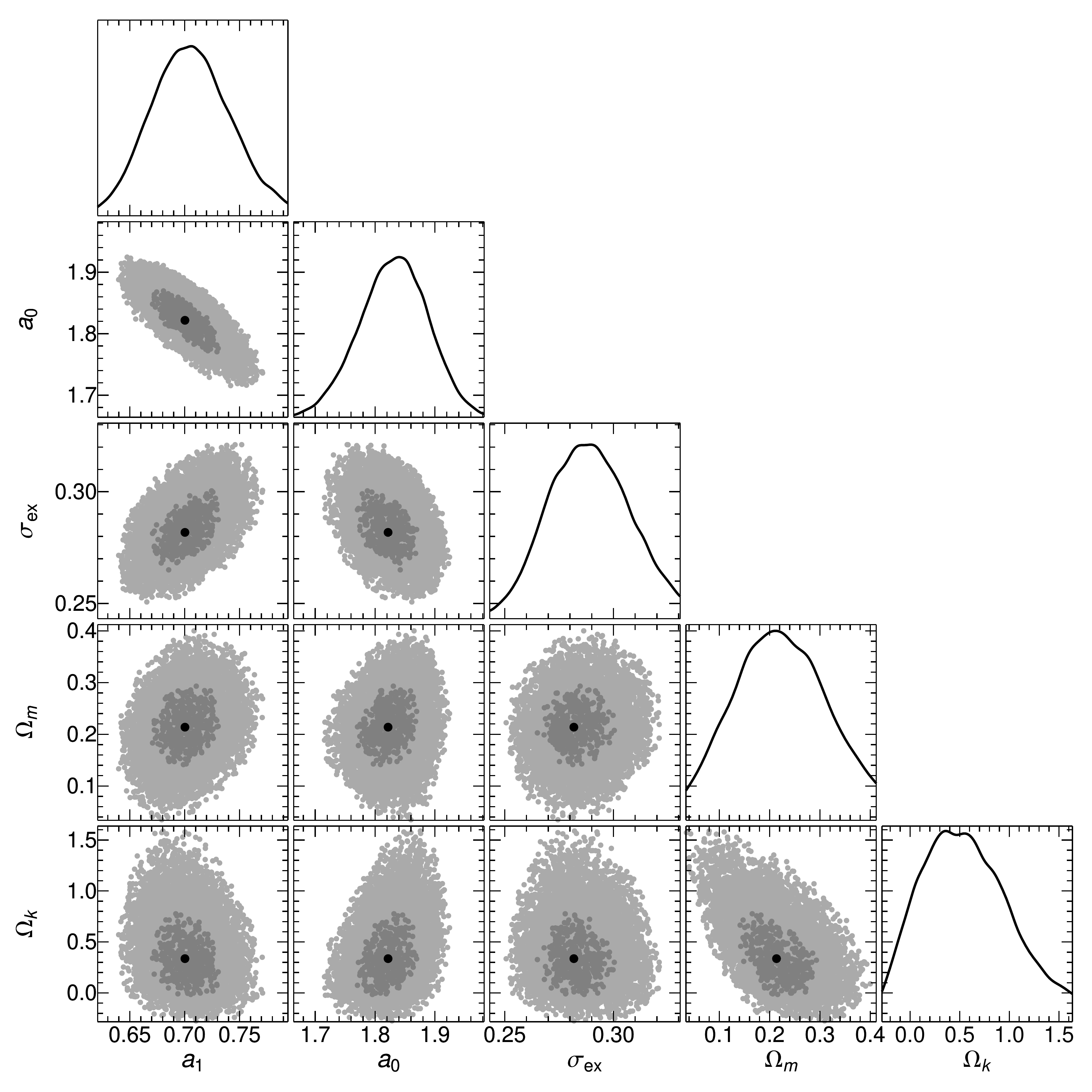}
\caption{Contour plots of the best-fit correlation and cosmological parameters (for a $\Lambda$CDM model with $\Omega_{\rm k}\neq0$): \textit{left} for $r=1$ and \textit{right} for $r\neq1$. Darker (lighter) regions mark the $1$--$\sigma$ ($2$--$\sigma$) confidence regions.}
\label{fig:Amati_contour}
\end{figure*}
\begin{figure*}
\centering
\includegraphics[width=0.49\hsize,clip]{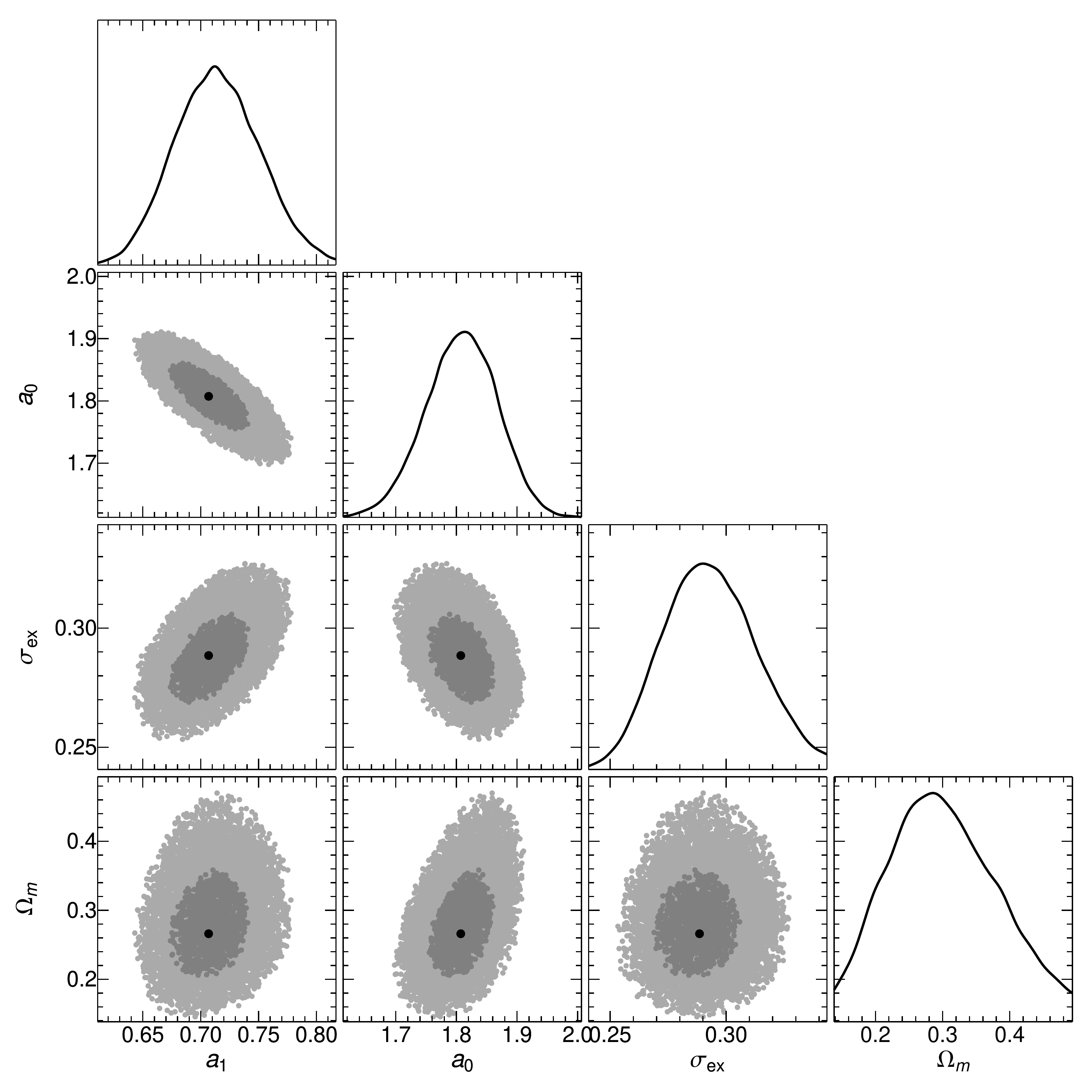}
\includegraphics[width=0.49\hsize,clip]{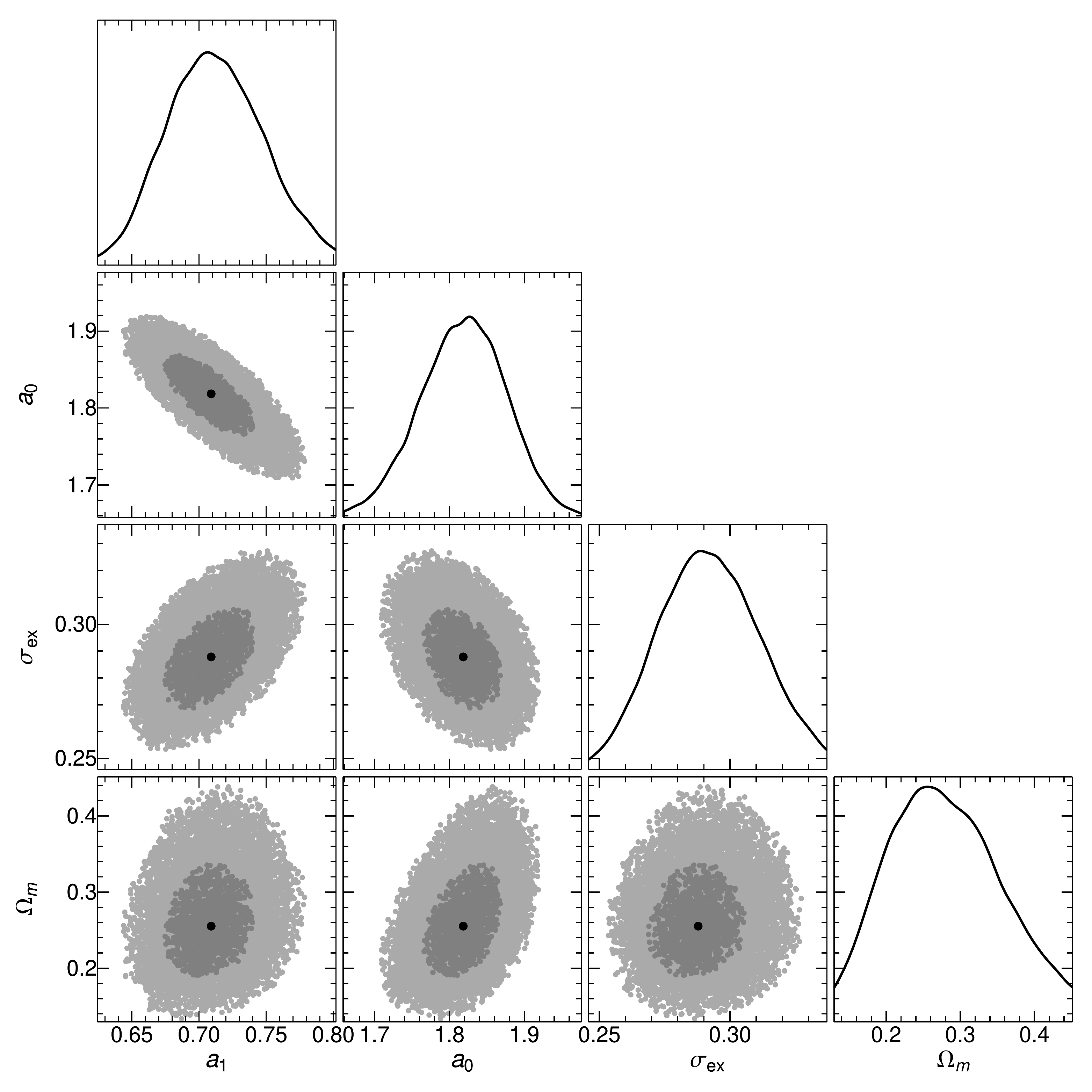}
\caption{The same as in Figs.~\ref{fig:Amati_contour} but for a $\Lambda$CDM model with $\Omega_{\rm k}=0$.}
\label{fig:Amati_flat_contour}
\end{figure*}


\section{Theoretical interpretation of numerical results}\label{sec:4}

The above-developed strategy definitely shows that the GRB calibration can be performed adopting intermediate data points with respect to previous efforts involving low-redshift catalogs. In other words, we demonstrate that the need of more low-redshift GRBs is not fully-motivated in order to get feasible strategies of GRB calibration that would fix the circularity issue. In this respect, we displayed in Figs. \ref{fig:Bez} suitable matching between theoretical and reconstructed curves. The case $r\neq1$ is slightly less predictive than the one with $r=1$, \textit{i.e.}, indicating that no particular deviations may occur as outputs of our MCMC analyses. This may be seen as a direct consequence of data reported in Table \ref{tab:BAO}, where $r$ is effectively close to unity. For both the cases, we highlight two normalized Hubble constants got from Eqs. \eqref{h0results}. It is intriguing to notice that the case $r\neq1$ provides a weakly larger value of $h_0$, albeit both cannot solve, in view of $2\sigma$ confidence level the Hubble tension today. Motivated by previous results toward this direction \citep{2021JCAP...09..042K,Khadka:2020hvb}, it seems evident that, even including GRBs into the computations of cosmic quantities, the tension remains unsolved. This conundrum could be therefore healed by switching to minimal extensions of the standard cosmological model \citep{Izzo2012c,Muccino:2020gqt,Luongo:2021nqh} or by additional terms within the Hilbert-Einstein action. 

Moving to our experimental findings, we notice from Figs. \ref{fig:Amati_contour} and \ref{fig:Amati_flat_contour} the remarkable fact the mass density is well constrained at the $1\sigma$ confidence level with respect to previous efforts making use of GRBs in the literature \citep{Cao:2021irf,Cao:2022wlg}. 
This certifies the goodness of intermediate calibrating techniques, demonstrating that our treatment is promising in refining cosmic outcomes by adopting GRBs. However, at $2\sigma$ confidence level the matter density appears weakly constrained, indicating that systematics may afflict the corresponding measurements. In general, $r=1$ measurements appear surprisingly better bounded than free $r$ parameters. In this case the values of $\Omega_m$ turn out to be closer to the Planck measurements in the case of spatially flat cosmology. For the sake of completeness, the same happens even for $r\neq1$, despite the non-flat case provides values of $\Omega_m$ completely outside any theoretical expectations. It is significant to stress that spatial curvature cannot be constrained properly even at $1\sigma$ confidence level, leaving open the task to fix it at intermediate and high redshifts. In particular, the values of spatial curvature is only slightly compatible with zero, being just approximately on the left error bar for both our measurements. All the remaining free parameters appear to be well constrained at both the $1\sigma$ and $2\sigma$ confidence levels.

In view of our findings, we conclude the background   cosmological model, namely the $\Lambda$CDM paradigm, is much more constrained by calibrating GRBs at intermediate redshifts. No significant expected departures in terms of dark energy are thus expected, but rather a plausible refinement of our fitting procedure, adding more data, could be useful to increase the values of matter densities here measured. A non-flat $\Lambda$CDM model is however debated and likely less probable than the flat case.

\section{Final outlooks and perspectives}
\label{sec:5}

In this paper, we proposed how to calibrate GRBs in a model-independent way, adopting since the very beginning intermediate-redshift data sets, instead of low-redshift catalogs as commonly performed in the literature. To do so, we employed the novel calibration technique that makes use of B\'ezier polynomials, applying its use to the $E_{\rm p}$--$E_{\rm iso}$ or \textit{Amati correlation}. In this respect, we adopted one of the best data set composed of $118$ GRBs and we worked out OHD and BAO surveys of data in conjunction by means of two B\'ezier parametric curves that allowed us to leave  \textit{a priori} free the spatial curvature. Hence,  \emph{standardizing} GRBs through the above technique, we tested the standard spatially flat $\Lambda$CDM model and its minimal extension adding a non-zero spatial curvature parameter, $\Omega_k$. 

Hence, MCMC analyses have been performed by means of the Metropolis algorithm in two main cases: $r=1$ and $r\neq1$, showing findings that were not perfectly compatible with the current expectations since they roughly differ from Planck results on $\Omega_k$. Thus, in our four cases, namely $r=1$ and $r\neq1$ with either zero or non-zero curvature, we provided tight bounds over the mass density, quite more similar to those found using standard candles. Further, we investigated the corresponding systematics plaguing the approach and we underlined that, although our outcomes looked more suitable than previous results got from the literature, the  $H_0$ tension was not addressed even with the use of GRBs. In analogy, the values of $\Omega_k$ were not bounded enough, but only slightly close to the zero value got from Planck measurements. 
Future works will focus on the use of refined B\'ezier model-independent curves and their use adopting further intermediate-redshift data points. Since  including intermediate data points would refined $\Omega_m$, we will see how the calibration can be made by \emph{both} low and intermediate data at the same time.


\section*{Acknowledgements}
The authors express their gratitude to Kuantay Boshkayev, Peter K. S. Dunsby and Francesco Pace for useful discussions. The work is financially supported by the Ministry of Education and Science of the Republic of Kazakhstan, Grant IRN AP08052311.


\bibliographystyle{mnras}

\begin{thebibliography}{}
\makeatletter
\relax
\def\mn@urlcharsother{\let\do\@makeother \do\$\do\&\do\#\do\^\do\_\do\%\do\~}
\def\mn@doi{\begingroup\mn@urlcharsother \@ifnextchar [ {\mn@doi@}
  {\mn@doi@[]}}
\def\mn@doi@[#1]#2{\def\@tempa{#1}\ifx\@tempa\@empty \href
  {http://dx.doi.org/#2} {doi:#2}\else \href {http://dx.doi.org/#2} {#1}\fi
  \endgroup}
\def\mn@eprint#1#2{\mn@eprint@#1:#2::\@nil}
\def\mn@eprint@arXiv#1{\href {http://arxiv.org/abs/#1} {{\tt arXiv:#1}}}
\def\mn@eprint@dblp#1{\href {http://dblp.uni-trier.de/rec/bibtex/#1.xml}
  {dblp:#1}}
\def\mn@eprint@#1:#2:#3:#4\@nil{\def\@tempa {#1}\def\@tempb {#2}\def\@tempc
  {#3}\ifx \@tempc \@empty \let \@tempc \@tempb \let \@tempb \@tempa \fi \ifx
  \@tempb \@empty \def\@tempb {arXiv}\fi \@ifundefined
  {mn@eprint@\@tempb}{\@tempb:\@tempc}{\expandafter \expandafter \csname
  mn@eprint@\@tempb\endcsname \expandafter{\@tempc}}}

\bibitem[\protect\citeauthoryear{{Alam} et~al.,}{{Alam}
  et~al.}{2017}]{2017MNRAS.470.2617A}
{Alam} S.,  et~al., 2017, \mn@doi [\mnras] {10.1093/mnras/stx721}, \href
  {https://ui.adsabs.harvard.edu/abs/2017MNRAS.470.2617A} {470, 2617}

\bibitem[\protect\citeauthoryear{{Amati} \& {Della Valle}}{{Amati} \& {Della
  Valle}}{2013}]{AmatiDellaValle2013}
{Amati} L.,  {Della Valle} M.,  2013, \mn@doi [International Journal of Modern
  Physics D] {10.1142/S0218271813300280}, \href
  {http://adsabs.harvard.edu/abs/2013IJMPD..2230028A} {22, 1330028}

\bibitem[\protect\citeauthoryear{{Amati} et~al.,}{{Amati}
  et~al.}{2002}]{Amati2002}
{Amati} L.,  et~al., 2002, \mn@doi [\aap] {10.1051/0004-6361:20020722}, \href {http://adsabs.harvard.edu/abs/2002A\%26A...390...81A} 
{390, 81}

\bibitem[\protect\citeauthoryear{{Amati}, {Guidorzi}, {Frontera}, {Della
  Valle}, {Finelli}, {Landi}  \& {Montanari}}{{Amati} et~al.}{2008}]{Amati2008}
{Amati} L.,  {Guidorzi} C.,  {Frontera} F.,  {Della Valle} M.,  {Finelli} F.,
  {Landi} R.,   {Montanari} E.,  2008, \mn@doi [\mnras]
  {10.1111/j.1365-2966.2008.13943.x}, \href
  {http://adsabs.harvard.edu/abs/2008MNRAS.391..577A} {391, 577}

\bibitem[\protect\citeauthoryear{{Amati}, {D'Agostino}, {Luongo}, {Muccino}  \&
  {Tantalo}}{{Amati} et~al.}{2019}]{2019MNRAS.486L..46A}
{Amati} L.,  {D'Agostino} R.,  {Luongo} O.,  {Muccino} M.,   {Tantalo} M.,
  2019, \mn@doi [\mnras] {10.1093/mnrasl/slz056}, \href
  {https://ui.adsabs.harvard.edu/abs/2019MNRAS.486L..46A} {486, L46}

\bibitem[\protect\citeauthoryear{{Anderson} et~al.,}{{Anderson}
  et~al.}{2014}]{2014MNRAS.441...24A}
{Anderson} L.,  et~al., 2014, \mn@doi [\mnras] {10.1093/mnras/stu523}, \href
  {https://ui.adsabs.harvard.edu/abs/2014MNRAS.441...24A} {441, 24}

\bibitem[\protect\citeauthoryear{{Arjona}, {Cardona}  \& {Nesseris}}{{Arjona}
  et~al.}{2019}]{2019PhRvD..99d3516A}
{Arjona} R.,  {Cardona} W.,   {Nesseris} S.,  2019, \mn@doi [\prd]
  {10.1103/PhysRevD.99.043516}, \href
  {https://ui.adsabs.harvard.edu/abs/2019PhRvD..99d3516A} {99, 043516}

\bibitem[\protect\citeauthoryear{{Ata} et~al.,}{{Ata}
  et~al.}{2018}]{2018MNRAS.473.4773A}
{Ata} M.,  et~al., 2018, \mn@doi [\mnras] {10.1093/mnras/stx2630}, \href
  {https://ui.adsabs.harvard.edu/abs/2018MNRAS.473.4773A} {473, 4773}

\bibitem[\protect\citeauthoryear{{Aubourg} et~al.,}{{Aubourg}
  et~al.}{2015}]{Aubourg15}
{Aubourg} {\'E}.,  et~al., 2015, \mn@doi [\prd] {10.1103/PhysRevD.92.123516},
  \href {http://adsabs.harvard.edu/abs/2015PhRvD..92l3516A} {92, 123516}

\bibitem[\protect\citeauthoryear{Aviles, Gruber, Luongo  \& Quevedo}{Aviles
  et~al.}{2012}]{Aviles:2012ay}
Aviles A.,  Gruber C.,  Luongo O.,   Quevedo H.,  2012, \mn@doi [Phys. Rev. D]
  {10.1103/PhysRevD.86.123516}, 86, 123516

\bibitem[\protect\citeauthoryear{{Bautista} et~al.,}{{Bautista}
  et~al.}{2018}]{2018ApJ...863..110B}
{Bautista} J.~E.,  et~al., 2018, \mn@doi [\apj] {10.3847/1538-4357/aacea5},
  \href {https://ui.adsabs.harvard.edu/abs/2018ApJ...863..110B} {863, 110}

\bibitem[\protect\citeauthoryear{{Belfiglio}, {Giamb{\`o}}  \&
  {Luongo}}{{Belfiglio} et~al.}{2022}]{mio2022}
{Belfiglio} A.,  {Giamb{\`o}} R.,   {Luongo} O.,  2022, arXiv e-prints, \href
  {https://ui.adsabs.harvard.edu/abs/2022arXiv220614158B} {p. arXiv:2206.14158}

\bibitem[\protect\citeauthoryear{{Bernardini}, {Margutti}, {Zaninoni}  \&
  {Chincarini}}{{Bernardini} et~al.}{2012}]{Bernardini2012}
{Bernardini} M.~G.,  {Margutti} R.,  {Zaninoni} E.,   {Chincarini} G.,  2012,
  \mn@doi [\mnras] {10.1111/j.1365-2966.2012.21487.x}, \href
  {http://adsabs.harvard.edu/abs/2012MNRAS.425.1199B} {425, 1199}

\bibitem[\protect\citeauthoryear{{Cao}, {Ryan}  \& {Ratra}}{{Cao}
  et~al.}{2021}]{2021MNRAS.504..300C}
{Cao} S.,  {Ryan} J.,   {Ratra} B.,  2021, \mn@doi [\mnras]
  {10.1093/mnras/stab942}, \href
  {https://ui.adsabs.harvard.edu/abs/2021MNRAS.504..300C} {504, 300}

\bibitem[\protect\citeauthoryear{Cao, Khadka  \& Ratra}{Cao
  et~al.}{2022a}]{Cao:2021irf}
Cao S.,  Khadka N.,   Ratra B.,  2022a, \mn@doi [Mon. Not. Roy. Astron. Soc.]
  {10.1093/mnras/stab3559}, 510, 2928

\bibitem[\protect\citeauthoryear{Cao, Dainotti  \& Ratra}{Cao
  et~al.}{2022b}]{Cao:2022wlg}
Cao S.,  Dainotti M.,   Ratra B.,  2022b, \mn@doi [Mon. Not. Roy. Astron. Soc.]
  {10.1093/mnras/stac517}, 512, 439

\bibitem[\protect\citeauthoryear{{Capozziello} \& {Izzo}}{{Capozziello} \&
  {Izzo}}{2008}]{CapozzielloIzzo2008}
{Capozziello} S.,  {Izzo} L.,  2008, \mn@doi [\aap]
  {10.1051/0004-6361:200810337}, \href
  {http://adsabs.harvard.edu/abs/2008A\%26A...490...31C} {490, 31}

\bibitem[\protect\citeauthoryear{{Capozziello}, {D'Agostino}  \&
  {Luongo}}{{Capozziello} et~al.}{2018}]{2018MNRAS.476.3924C}
{Capozziello} S.,  {D'Agostino} R.,   {Luongo} O.,  2018, \mn@doi [\mnras]
  {10.1093/mnras/sty422}, \href
  {https://ui.adsabs.harvard.edu/abs/2018MNRAS.476.3924C} {476, 3924}

\bibitem[\protect\citeauthoryear{{Capozziello}, {D'Agostino}  \&
  {Luongo}}{{Capozziello} et~al.}{2019}]{2019IJMPD..2830016C}
{Capozziello} S.,  {D'Agostino} R.,   {Luongo} O.,  2019, \mn@doi
  [International Journal of Modern Physics D] {10.1142/S0218271819300167},
  \href {https://ui.adsabs.harvard.edu/abs/2019IJMPD..2830016C} {28, 1930016}

\bibitem[\protect\citeauthoryear{{Capozziello}, {D'Agostino}  \&
  {Luongo}}{{Capozziello} et~al.}{2020}]{2020arXiv200309341C}
{Capozziello} S.,  {D'Agostino} R.,   {Luongo} O.,  2020, arXiv e-prints, \href
  {https://ui.adsabs.harvard.edu/abs/2020arXiv200309341C} {p. arXiv:2003.09341}

\bibitem[\protect\citeauthoryear{{Carter}, {Beutler}, {Percival}, {Blake},
  {Koda}  \& {Ross}}{{Carter} et~al.}{2018}]{2018MNRAS.481.2371C}
{Carter} P.,  {Beutler} F.,  {Percival} W.~J.,  {Blake} C.,  {Koda} J.,
  {Ross} A.~J.,  2018, \mn@doi [\mnras] {10.1093/mnras/sty2405}, \href
  {https://ui.adsabs.harvard.edu/abs/2018MNRAS.481.2371C} {481, 2371}

\bibitem[\protect\citeauthoryear{{Cucchiara} et~al.,}{{Cucchiara}
  et~al.}{2011}]{Cucchiara2011}
{Cucchiara} A.,  et~al., 2011, \mn@doi [\apj] {10.1088/0004-637X/736/1/7},
  \href {http://adsabs.harvard.edu/abs/2011ApJ...736....7C} {736, 7}

\bibitem[\protect\citeauthoryear{{D'Agostini}}{{D'Agostini}}{2005}]{Dago2005}
{D'Agostini} G.,  2005, ArXiv Physics e-prints, \href
  {http://adsabs.harvard.edu/abs/2005physics..11182D} {}

\bibitem[\protect\citeauthoryear{{D'Agostino}, {Luongo}  \&
  {Muccino}}{{D'Agostino} et~al.}{2022}]{2022arXiv220402190D}
{D'Agostino} R.,  {Luongo} O.,   {Muccino} M.,  2022, arXiv e-prints, \href
  {https://ui.adsabs.harvard.edu/abs/2022arXiv220402190D} {p. arXiv:2204.02190}

\bibitem[\protect\citeauthoryear{Dainotti \& Amati}{Dainotti \&
  Amati}{2018}]{Dainotti18}
Dainotti M.~G.,  Amati L.,  2018, Publications of the Astronomical Society of
  the Pacific, 130, 051001

\bibitem[\protect\citeauthoryear{{Dainotti}, {Cardone}  \&
  {Capozziello}}{{Dainotti} et~al.}{2008}]{Dainotti2008}
{Dainotti} M.~G.,  {Cardone} V.~F.,   {Capozziello} S.,  2008, \mn@doi [\mnras]
  {10.1111/j.1745-3933.2008.00560.x}, \href
  {http://esoads.eso.org/abs/2008MNRAS.391L..79D} {391, L79}

\bibitem[\protect\citeauthoryear{{Demianski}, {Piedipalumbo}, {Sawant}  \&
  {Amati}}{{Demianski} et~al.}{2017a}]{Demianski17a}
{Demianski} M.,  {Piedipalumbo} E.,  {Sawant} D.,   {Amati} L.,  2017a, \mn@doi
  [\aap] {10.1051/0004-6361/201628909}, \href
  {http://adsabs.harvard.edu/abs/2017A\%26A...598A.112D} {598, A112}

\bibitem[\protect\citeauthoryear{{Demianski}, {Piedipalumbo}, {Sawant}  \&
  {Amati}}{{Demianski} et~al.}{2017b}]{Demianski17b}
{Demianski} M.,  {Piedipalumbo} E.,  {Sawant} D.,   {Amati} L.,  2017b, \mn@doi
  [\aap] {10.1051/0004-6361/201628911}, \href
  {http://adsabs.harvard.edu/abs/2017A\%26A...598A.113D} {598, A113}

\bibitem[\protect\citeauthoryear{{Di Valentino} et~al.,}{{Di Valentino}
  et~al.}{2021}]{2021CQGra..38o3001D}
{Di Valentino} E.,  et~al., 2021, \mn@doi [Classical and Quantum Gravity]
  {10.1088/1361-6382/ac086d}, \href
  {https://ui.adsabs.harvard.edu/abs/2021CQGra..38o3001D} {38, 153001}

\bibitem[\protect\citeauthoryear{Dunsby \& Luongo}{Dunsby \&
  Luongo}{2016}]{Dunsby:2015ers}
Dunsby P. K.~S.,  Luongo O.,  2016, \mn@doi [Int. J. Geom. Meth. Mod. Phys.]
  {10.1142/S0219887816300026}, 13, 1630002

\bibitem[\protect\citeauthoryear{{Ghirlanda}, {Ghisellini}, {Lazzati}  \&
  {Firmani}}{{Ghirlanda} et~al.}{2004}]{Ghirlanda04}
{Ghirlanda} G.,  {Ghisellini} G.,  {Lazzati} D.,   {Firmani} C.,  2004, \mn@doi
  [\apjl] {10.1086/424915}, \href
  {http://adsabs.harvard.edu/abs/2004ApJ...613L..13G} {613, L13}

\bibitem[\protect\citeauthoryear{{Goobar} \& {Perlmutter}}{{Goobar} \&
  {Perlmutter}}{1995}]{Goobar1995}
{Goobar} A.,  {Perlmutter} S.,  1995, \mn@doi [\apj] {10.1086/176113}, \href
  {http://adsabs.harvard.edu/abs/1995ApJ...450...14G} {450, 14}

\bibitem[\protect\citeauthoryear{{Izzo}, {Luongo}  \& {Capozziello}}{{Izzo}
  et~al.}{2012}]{Izzo2012c}
{Izzo} L.,  {Luongo} O.,   {Capozziello} S.,  2012, Memorie della Societa
  Astronomica Italiana Supplementi, \href
  {http://adsabs.harvard.edu/abs/2012MSAIS..19...37I} {19, 37}

\bibitem[\protect\citeauthoryear{{Izzo}, {Muccino}, {Zaninoni}, {Amati}  \&
  {Della Valle}}{{Izzo} et~al.}{2015}]{Izzo2015}
{Izzo} L.,  {Muccino} M.,  {Zaninoni} E.,  {Amati} L.,   {Della Valle} M.,
  2015, \mn@doi [\aap] {10.1051/0004-6361/201526461}, \href
  {http://adsabs.harvard.edu/abs/2015A\%26A...582A.115I} {582, A115}

\bibitem[\protect\citeauthoryear{{Kazin} et~al.,}{{Kazin}
  et~al.}{2014}]{2014MNRAS.441.3524K}
{Kazin} E.~A.,  et~al., 2014, \mn@doi [\mnras] {10.1093/mnras/stu778}, \href
  {https://ui.adsabs.harvard.edu/abs/2014MNRAS.441.3524K} {441, 3524}

\bibitem[\protect\citeauthoryear{Khadka \& Ratra}{Khadka \&
  Ratra}{2020}]{Khadka:2020hvb}
Khadka N.,  Ratra B.,  2020, \mn@doi [Mon. Not. Roy. Astron. Soc.]
  {10.1093/mnras/staa2779}, 499, 391

\bibitem[\protect\citeauthoryear{{Khadka}, {Luongo}, {Muccino}  \&
  {Ratra}}{{Khadka} et~al.}{2021}]{2021JCAP...09..042K}
{Khadka} N.,  {Luongo} O.,  {Muccino} M.,   {Ratra} B.,  2021, \mn@doi [\jcap]
  {10.1088/1475-7516/2021/09/042}, \href
  {https://ui.adsabs.harvard.edu/abs/2021JCAP...09..042K} {2021, 042}

\bibitem[\protect\citeauthoryear{{Luongo} \& {Muccino}}{{Luongo} \&
  {Muccino}}{2018}]{nostro}
{Luongo} O.,  {Muccino} M.,  2018, \mn@doi [\prd] {10.1103/PhysRevD.98.103520},
  \href {https://ui.adsabs.harvard.edu/abs/2018PhRvD..98j3520L} {98, 103520}

\bibitem[\protect\citeauthoryear{{Luongo} \& {Muccino}}{{Luongo} \&
  {Muccino}}{2021a}]{2021Galax...9...77L}
{Luongo} O.,  {Muccino} M.,  2021a, \mn@doi [Galaxies]
  {10.3390/galaxies9040077}, \href
  {https://ui.adsabs.harvard.edu/abs/2021Galax...9...77L} {9, 77}

\bibitem[\protect\citeauthoryear{{Luongo} \& {Muccino}}{{Luongo} \&
  {Muccino}}{2021b}]{LM2020}
{Luongo} O.,  {Muccino} M.,  2021b, \mn@doi [\mnras] {10.1093/mnras/stab795},
  \href {https://ui.adsabs.harvard.edu/abs/2021MNRAS.503.4581L} {503, 4581}

\bibitem[\protect\citeauthoryear{Luongo, Muccino, Colg\'ain, Sheikh-Jabbari  \&
  Yin}{Luongo et~al.}{2022}]{Luongo:2021nqh}
Luongo O.,  Muccino M.,  Colg\'ain E.~O.,  Sheikh-Jabbari M.~M.,   Yin L.,
  2022, \mn@doi [Phys. Rev. D] {10.1103/PhysRevD.105.103510}, 105, 103510

\bibitem[\protect\citeauthoryear{Muccino, Izzo, Luongo, Boshkayev, Amati,
  Della~Valle, Pisani  \& Zaninoni}{Muccino et~al.}{2021}]{Muccino:2020gqt}
Muccino M.,  Izzo L.,  Luongo O.,  Boshkayev K.,  Amati L.,  Della~Valle M.,
  Pisani G.~B.,   Zaninoni E.,  2021, \mn@doi [Astrophys. J.]
  {10.3847/1538-4357/abd254}, 908, 181

\bibitem[\protect\citeauthoryear{{Ooba}, {Ratra}  \& {Sugiyama}}{{Ooba}
  et~al.}{2018}]{2018ApJ...864...80O}
{Ooba} J.,  {Ratra} B.,   {Sugiyama} N.,  2018, \mn@doi [\apj]
  {10.3847/1538-4357/aad633}, \href
  {https://ui.adsabs.harvard.edu/\#abs/2018ApJ...864...80O} {864, 80}

\bibitem[\protect\citeauthoryear{{Padmanabhan}, {Xu}, {Eisenstein}, {Scalzo},
  {Cuesta}, {Mehta}  \& {Kazin}}{{Padmanabhan}
  et~al.}{2012}]{2012MNRAS.427.2132P}
{Padmanabhan} N.,  {Xu} X.,  {Eisenstein} D.~J.,  {Scalzo} R.,  {Cuesta} A.~J.,
   {Mehta} K.~T.,   {Kazin} E.,  2012, \mn@doi [\mnras]
  {10.1111/j.1365-2966.2012.21888.x}, \href
  {https://ui.adsabs.harvard.edu/abs/2012MNRAS.427.2132P} {427, 2132}

\bibitem[\protect\citeauthoryear{{Percival} et~al.,}{{Percival}
  et~al.}{2010}]{Percival10}
{Percival} W.~J.,  et~al., 2010, \mn@doi [\mnras]
  {10.1111/j.1365-2966.2009.15812.x}, \href
  {http://adsabs.harvard.edu/abs/2010MNRAS.401.2148P} {401, 2148}

\bibitem[\protect\citeauthoryear{{Perivolaropoulos} \&
  {Skara}}{{Perivolaropoulos} \& {Skara}}{2021}]{2021arXiv210505208P}
{Perivolaropoulos} L.,  {Skara} F.,  2021, arXiv e-prints, \href
  {https://ui.adsabs.harvard.edu/abs/2021arXiv210505208P} {p. arXiv:2105.05208}

\bibitem[\protect\citeauthoryear{{Planck Collaboration}}{{Planck
  Collaboration}}{2020}]{Planck2018}
{Planck Collaboration} 2020, \mn@doi [\aap] {10.1051/0004-6361/201833910},
  \href {https://ui.adsabs.harvard.edu/abs/2020A&A...641A...6P} {641, A6}

\bibitem[\protect\citeauthoryear{{Ratra} \& {Peebles}}{{Ratra} \&
  {Peebles}}{1988}]{Ratra1988}
{Ratra} B.,  {Peebles} P.~J.~E.,  1988, \mn@doi [\prd]
  {10.1103/PhysRevD.37.3406}, \href
  {http://adsabs.harvard.edu/abs/1988PhRvD..37.3406R} {37, 3406}

\bibitem[\protect\citeauthoryear{{Riess} et~al.,}{{Riess}
  et~al.}{2018}]{2018ApJ...853..126R}
{Riess} A.~G.,  et~al., 2018, \mn@doi [\apj] {10.3847/1538-4357/aaa5a9}, \href
  {https://ui.adsabs.harvard.edu/abs/2018ApJ...853..126R} {853, 126}

\bibitem[\protect\citeauthoryear{{Rodney} et~al.,}{{Rodney}
  et~al.}{2015}]{Rodney2015}
{Rodney} S.~A.,  et~al., 2015, \mn@doi [\aj] {10.1088/0004-6256/150/5/156},
  \href {http://adsabs.harvard.edu/abs/2015AJ....150..156R} {150, 156}

\bibitem[\protect\citeauthoryear{{Salvaterra} et~al.,}{{Salvaterra}
  et~al.}{2009}]{Salvaterra2009}
{Salvaterra} R.,  et~al., 2009, \mn@doi [\nat] {10.1038/nature08445}, \href
  {http://adsabs.harvard.edu/abs/2009Natur.461.1258S} {461, 1258}

\bibitem[\protect\citeauthoryear{{Schaefer}}{{Schaefer}}{2007}]{Schaefer2007}
{Schaefer} B.~E.,  2007, \mn@doi [\apj] {10.1086/511742}, \href
  {http://adsabs.harvard.edu/abs/2007ApJ...660...16S} {660, 16}

\bibitem[\protect\citeauthoryear{{Scolnic} et~al.,}{{Scolnic}
  et~al.}{2018}]{2018ApJ...859..101S}
{Scolnic} D.~M.,  et~al., 2018, \mn@doi [\apj] {10.3847/1538-4357/aab9bb},
  \href {https://ui.adsabs.harvard.edu/abs/2018ApJ...859..101S} {859, 101}

\bibitem[\protect\citeauthoryear{Sotiriou \& Faraoni}{Sotiriou \&
  Faraoni}{2010}]{Sotiriou:2008rp}
Sotiriou T.~P.,  Faraoni V.,  2010, \mn@doi [Rev. Mod. Phys.]
  {10.1103/RevModPhys.82.451}, 82, 451

\bibitem[\protect\citeauthoryear{{Tanvir} et~al.,}{{Tanvir}
  et~al.}{2009}]{Tanvir2009}
{Tanvir} N.~R.,  et~al., 2009, \mn@doi [\nat] {10.1038/nature08459}, \href
  {http://adsabs.harvard.edu/abs/2009Natur.461.1254T} {461, 1254}

\bibitem[\protect\citeauthoryear{{Tsujikawa}}{{Tsujikawa}}{2013}]{Tsujikawa2013}
{Tsujikawa} S.,  2013, \mn@doi [Classical and Quantum Gravity]
  {10.1088/0264-9381/30/21/214003}, \href
  {http://adsabs.harvard.edu/abs/2013CQGra..30u4003T} {30, 214003}

\bibitem[\protect\citeauthoryear{{Wei}, {Wu}, {Melia}, {Wei}  \& {Feng}}{{Wei}
  et~al.}{2014}]{Wei2014}
{Wei} J.-J.,  {Wu} X.-F.,  {Melia} F.,  {Wei} D.-M.,   {Feng} L.-L.,  2014,
  \mn@doi [\mnras] {10.1093/mnras/stu166}, \href
  {http://adsabs.harvard.edu/abs/2014MNRAS.439.3329W} {439, 3329}

\bibitem[\protect\citeauthoryear{{du Mas des Bourboux} et~al.,}{{du Mas des
  Bourboux} et~al.}{2020}]{2020ApJ...901..153D}
{du Mas des Bourboux} H.,  et~al., 2020, \mn@doi [\apj]
  {10.3847/1538-4357/abb085}, \href
  {https://ui.adsabs.harvard.edu/abs/2020ApJ...901..153D} {901, 153}

\makeatother
\end{thebibliography}

\end{document}